\documentclass[aps,prb,twocolumn,groupedaddress,showpacs,floatfix]{revtex4-1}

\usepackage{graphicx}
\usepackage{subfigure}
\usepackage{amsmath}
\usepackage{amssymb}
\graphicspath{{./}{figures/}}

\newcommand{\LaAs}{LaFeAsO}
\newcommand{\Ba}{BaFe$_2$As$_2$}
\newcommand{\Sr}{SrFe$_2$As$_2$}
\newcommand{\Eu}{EuFe$_2$As$_2$}

\def\muB{\ensuremath{\mu_{\rm B}}}
\def\vec#1{{\boldsymbol{#1}}}
\def\bra#1{\ensuremath{\langle{#1}\vert}}
\def\ket#1{\ensuremath{\vert{#1}\rangle}}
\DeclareMathOperator{\re}{Re}
\DeclareMathOperator{\im}{Im}

\begin{document}

%Title of paper
\title{Analysis of spin density wave conductivity spectra of iron pnictides in the framework of density functional theory}

\author{Johannes Ferber}
\email{ferber@itp.uni-frankfurt.de}

\author{Yu-Zhong Zhang}
%\email{yzhang@itp.uni-frankfurt.de}

\author{Harald O. Jeschke}
%\email{jeschke@itp.uni-frankfurt.de}

\author{Roser Valent\'\i}
%\email{valenti@itp.uni-frankfurt.de}

\affiliation{Institut f\"ur Theoretische Physik, Goethe-Universit\"at Frankfurt, Max-von-Laue-Strasse 1, 60438 Frankfurt/Main, Germany}

\date{\today}

\begin{abstract}
  The optical conductivity of {\LaAs}, \Ba, {\Sr}, and {\Eu} in the
  spin-density wave (SDW) state is investigated within density
  functional theory (DFT) in the framework of spin-polarized
  generalized gradient approximation (GGA) and GGA+U. We find a strong
  dependence of the optical features on the Fe magnetic moments. In
  order to recover the small Fe magnetic moments observed
  experimentally, GGA+$U_{\rm eff}$ with a suitable choice of negative
  on-site interaction $U_{\rm eff}=U-J$ was considered. Such an
  approach may be justified in terms of an overscreening which induces
  a relatively small $U$ compared to the Hund's rule coupling $J$, as
  well as a strong Holstein-like electron-phonon
  interaction. Moreover, reminiscent of the fact that GGA+$U_{\rm
    eff}$ with a positive $U_{\rm eff}$ is a simple approximation for reproducing a gap with correct amplitude
in correlated insulators, a negative $U_{\rm eff}$
  can also be understood as a way to suppress magnetism and mimic the effects of quantum fluctuations ignored in DFT calculations.
With these considerations, the resulting optical spectra reproduce the
SDW gap and a number of experimentally observed features related to
the antiferromagnetic order. We find electronic contributions to
excitations that so far have been attributed to purely phononic
modes. Also, an orbital resolved analysis of the optical conductivity
reveals significant contributions from all Fe~$3d$ orbitals. Finally,
we observe that there is an important renormalization of kinetic
energy in these SDW metals, implying that the effects of correlations
cannot be neglected.
\end{abstract}


\maketitle

\section{Introduction}

In several families of the iron pnictides, high temperature
superconductivity (SC) emerges in close proximity of an
antiferromagnetic (AF) ground state with stripe-type
order.~\cite{Ishida2009} The AF transition is induced by a
spin-density wave (SDW) instability below a critical temperature
$T_{\rm SDW}$, which is either preceded or coincidental with a
structural transition from a tetragonal phase to an orthorhombic
phase.~\cite{Lynn2009} Upon doping~\cite{Kamihara2008} or
application of pressure~\cite{Torikachvili2008}, the AF order is
suppressed and superconductivity emerges. Like lattice vibrations, AF
fluctuations are capable of producing an attractive interaction
necessary for the creation of Cooper pairs. It is therefore widely
believed that the AF fluctuations drive the superconducting
instability in these systems. As for the nature of the AF state, the
iron pnictides show signatures of both electron itinerancy and local
magnetism. Some
experimental~\cite{Hu2008,Hsieh2008,Yi2009,Fink2009,Kimber2009,Thirupathaiah2010,Chen2010}
and theoretical works based on density functional theory
(DFT)~\cite{Mazin2008,Dong2008,Cvetkovic2009,Opahle2009,Zhang2009,Yaresko2009,Zhang2010}
and dynamical mean-field theory (DMFT)~\cite{Lee2009} as well as a
combination of these two methods~\cite{Aichhorn2009,Skornyakov2009}
favor an itinerant scenario with rather moderate correlation strength,
as reflected by the metallic nature of the compounds. On the other
hand, other authors point out the effects of strong
correlations,~\cite{Yang2009,Si2008,Haule2008,Yildirim2008,Han2009,Ma2009}
like the renormalization of the kinetic energy of the
electrons.~\cite{Qazilbash2009}

Many features of the electronic structure of the iron pnictides are
directly reflected in their optical properties: the low-frequency
region of the conductivity spectrum is governed by the itinerant
carrier contribution and directly shows the effect of correlations in
the area under the Drude region which is proportional to the
electron's kinetic energy; the infrared regime above the Drude peak is
dominated by gap features induced by either the SDW gap or the
superconducting gap; finally, the visible part of the spectrum
reflects the band structure in the normal state.  Consequently, a
number of experimental studies have been performed on the optical
properties of the iron pnictides, in the normal state, the SDW state,
as well as in the SC
state.~\cite{Hu2008,Drechsler2008,Nakajima2010,Chen2010,Wu2010,Wu2009a,Gorshunov2010,Lucarelli2010}
The SDW state is characterized by (i) the appearance of a peak in the
optical conductivity at the SDW gap frequency, (ii) an anisotropic dc
response and (iii) an almost isotropic response in the infrared and
optical region of the spectrum. On the other hand, while several
theoretical works based on local density approximation
(LDA)~\cite{Qazilbash2009} and LDA+DMFT~\cite{Haule2008,Laad2009} have
been done on the paramagnetic phase, there is still a lack of DFT
calculations for the optical conductivity in the SDW state.

In this work we report optical studies in the framework of density
functional theory on four iron pnictides in the SDW state, namely the
1111 compound {\LaAs}, and the 122 compounds \textit{AE}Fe$_2$As$_2$
(\textit{AE}=Ba,Sr,Eu). This provides an insight into the microscopic
origin of the optical features in the SDW state and allows for an
assessment of DFT regarding its applicability to the iron pnictides.

\section{Computational Details}
\label{section:computational_details}

We performed electronic structure calculations with the full potential
linearized augmented plane wave (FLAPW) method as implemented in
\texttt{WIEN2K}.\cite{Blaha2001} The self-consistency cycle employed
2048 $k$ points in the full Brillouin zone (FBZ) using the generalized
gradient approximation (GGA) in the Perdew-Burke-Ernzerhof variant for
the exchange correlation potential;~\cite{Perdew1996} the optical
properties were evaluated with 16384 $k$ points in the FBZ, the number of
$k$ points in the irreducible Brillouin zone depends on the symmetries
of the respective material.  Experimental lattice parameters and
atomic positions were used, from Refs.~\onlinecite{Nomura2008,
  Huang2008, Alireza2009, Tegel2008} for {\LaAs}, \Ba, {\Sr}, and
{\Eu}, respectively. All calculations were performed in the scalar
relativistic approximation. For the optical properties, the
\texttt{optics}~\cite{Ambrosch-Draxl2006} code package in
\texttt{WIEN2K} was modified to allow for an orbital
character resolved analysis.

We are working in the framework of 'GGA+U' where '$U$'$(\equiv
U_{\rm eff}=U-J)$ describes the competition between the (spherically averaged) on-site
Coulomb interaction $U$ and the (spherically averaged) on-site exchange coupling $J$
(within the Fe~$3d$ subshell for the iron pnictides). In this
context, the atomic limit double counting correction according to
Refs.~\onlinecite{Anisimov1993,Dudarev1998} was applied. With this double counting correction,
the expression for the correction to the GGA functional for the orbital $m$ reads
\begin{equation}
 \frac{U-J}{2}\sum_{\sigma} n_{m\sigma}(1-n_{m\sigma}),
\label{eq:u_correction}
\end{equation}
where $n_{m\sigma}$ is the spin-projected occupation in orbital $m$.

In order to allow for the stripe-type AF order, we consider a doubled
($\sqrt{2}\times\sqrt{2}\times 1$) unit cell with AF order along the
$a$ axis of the supercell and ferromagnetic (FM) arrangement along the
$b$ axis ({\it i.e.} the supercell is rotated 45$^{\circ}$ with
respect to the original unit cell), as observed experimentally. In the
following, the orbital characters and dielectric tensor components are
labelled with respect to the coordinate system of this
supercell. Spin-polarized calculations with AF order are labelled with
'GGA(AF)' ('GGA+U(AF)', respectively).

The linear optical response of the electronic
system~\footnote{Phononic excitations are not considered in this work}
is calculated in the random phase approximation (RPA), with the
Kohn-Sham orbitals mimicking the bare electronic states of the
electron-hole excitations in the RPA formalism. In this framework, the
complex-valued dielectric function $\epsilon$ is calculated as
\begin{equation}\begin{split}
\epsilon(\omega)=1-&\lim_{q\rightarrow 0} \frac{4\pi e^2}{\Omega_c |\vec{q}|^2} \sum_{n,n',\vec{k}}\frac{f_0(E_{n',\vec{k}+\vec{q}})-f_0(E_{n,\vec{k}})}{E_{n',\vec{k}+\vec{q}}-E_{n,\vec{k}}-\hbar\omega}\\
&\times|M_{n,n'}(\vec{k},\vec{q})|^2, \label{eq:epsilon}
\end{split}\end{equation}
where $\Omega_c$ is the volume of the unit cell, $\vec{q}$ the
wavevector of the incoming light, $f_0$ the Fermi distribution, and
$\{n,k\}$ labels the Kohn-Sham (KS) orbital in band $n$ with crystal
momentum $k$. The transition matrix element is given by
$M_{n,n'}(\vec{k},\vec{q})=\bra{n,\vec{k}}e^{-i\vec{q}\vec{r}}\ket{n',\vec{k}+\vec{q}}$.
As expressed by $\lim_{q\rightarrow 0}$, only direct transitions
$\vec{k}_f=\vec{k}_i$ are considered here since even for high
energies, the wavelength of the light is large compared to the lattice
dimensions. In this limit, perturbation theory for
$\vec{k}\cdot\vec{p}$ can be applied ($\vec{p}$ being the momentum
operator) in order to obtain expressions for $\ket{n,\vec{k}+\vec{q}}$
and $E_{n,\vec{k}+\vec{q}}$ in terms of $\ket{n,\vec{k}}$ and
$E_{n,\vec{k}}$ within the dipole approximation. This yields for the
dielectric tensor~\cite{Ambrosch-Draxl2006}
\begin{equation}\begin{split}
&\epsilon_{ij}(\omega)=1+\frac{4\pi \hbar^2 e^2}{\Omega_c m^2 \omega^2} \sum_{n,\vec{k}} \left(\frac{\partial f_0}{\partial E}\right)_{E_{\rm F}} p_{i;n,n,\vec{k}} \; p_{j;n,n,\vec{k}}\\
&-\frac{4\pi \hbar^2 e^2}{\Omega_c m^2} \sum_{\vec{k}}
\sum_{v,c}\frac{p_{i;c,v,\vec{k}} \;
p_{j;c,v,\vec{k}}}{(E_{c,\vec{k}}\!-\!E_{v,\vec{k}}\!-\!\hbar\omega)(E_{c,\vec{k}}\!-\!E_{v,\vec{k}})^2},
\label{eq:epsilon_tensor}
\end{split}\end{equation}
where the second and third term describe the intraband and interband
contributions, respectively.  Since $T=0$ is considered, the intraband
contribution is restricted to states at the Fermi energy $E_{\rm F}$,
whereas the band indices $v$($c$) run over all occupied (empty)
states. $p_i$ is the matrix element of the momentum operator along the
electric field polarization of the incoming light,
$p_{i;n,n',\vec{k}}=\bra{n,\vec{k}}p_i\ket{n',\vec{k}}$. Spin-orbit
coupling is not taken into account in our calculations, and therefore
all off-diagonal elements $i\neq j$ vanish due to the orthorhombic
symmetry of the crystals.

Turning the sum in Eq.~(\ref{eq:epsilon_tensor}) into an integral
over the first Brillouin zone and using the Dirac representation
$\frac{1}{x \pm i \eta}=\mathcal{P} \frac{1}{x} \mp i \pi
\delta(x)$, the imaginary part of the interband contributions to the
dielectric tensor components can be expressed as
\begin{equation} \im
\epsilon_{ii}^{\rm inter}(\omega)=\frac{\hbar^2 e^2}{\pi m^2
\omega^2} \sum_{v,c} \int_{\vec{k}} |p_{i;c,v,\vec{k}}|^2
\delta(E_{c,\vec{k}}-E_{v,\vec{k}}-\hbar\omega).
\label{eq:epsilon_interband}
\end{equation}

In this work, we focus on the analysis of the real part of the optical
conductivity,
\begin{equation}
\re \sigma_{ij}(\omega)=\frac{\omega}{4 \pi} \im \epsilon_{ij}(\omega).
\end{equation}
For metals, the intraband contribution has the well-known Drude shape
\begin{equation}
\sigma_D=\frac{\Gamma \omega_p^2}{4\pi(\omega^2+\Gamma^2)}\;,
\end{equation}
where the plasma frequency $\omega_p$ determines the spectral weight
of the Drude peak via $\int_0^\infty d\omega\,\sigma_D(\omega)
=\frac{\omega_p}{8}$, whereas its width is given by the carrier
scattering rate $\Gamma=1/\tau$ with the lifetime $\tau$.

The presence of the Drude peak depends on the metallicity of the
material under consideration and in general needs to be confirmed by
an explicit calculation of the intraband transitions. However, as only
direct transitions are considered here, the intraband part of
Eq.~(\ref{eq:epsilon_tensor}) is singular at $\omega=0$ and vanishes
at finite frequencies, both for the dielectric function as well as for
the conductivity. Instead, the squared plasma frequency for band $n$
and spin $\sigma$ is defined as
\begin{equation}
\omega_p^2(n,\sigma)=\frac{\hbar^2 e^2}{\pi m^2} \int_{\vec{k}} |\bra{n,\vec{k},\sigma}p\ket{n,\vec{k},\sigma}|^2 \delta(E_{n,\vec{k},\sigma}-E_{\rm F})
\label{eq:plasmaf}
\end{equation}
with the total squared plasma frequency
$\omega_p^2=\sum_{n,\sigma}\omega_p^2(n,\sigma)$, and the Drude peak
is considered as an approximation to the intraband conductivity, {\it
  i.e.} we consider only one overall Drude peak, despite the
possibility of two or more Drude peaks due to multiple bands crossing
the Fermi surface. For the iron pnictides, this is justified as they
show metallic behaviour also below the SDW transition. Consequently,
only the carrier scattering rate is left as free parameter which
cannot be determined within a DFT approach. For our purposes, $\Gamma$
is chosen close to experimental values~\cite{Hu2008} in the SDW
state. Given these considerations, all relevant
information about the Drude peaks is contained in the plasma frequency $\omega_p$.

\section{Results}

\subsection{Density of States}

Preliminary to the analysis of the optical properties of the iron
pnictides, we first start with the discussion of the density of states
(DOS) since the dielectric function is essentially the joint density
of states (jDOS) weighted with the transition matrix element (see
Eq.~(\ref{eq:epsilon_interband})). In particular, we find that the
antiferromagnetic order and the magnitude of the magnetic moments
crucially affect the optical properties which is reflected by the
(partial) opening of a gap in the DOS.

\begin{figure}
\includegraphics[width=\columnwidth]{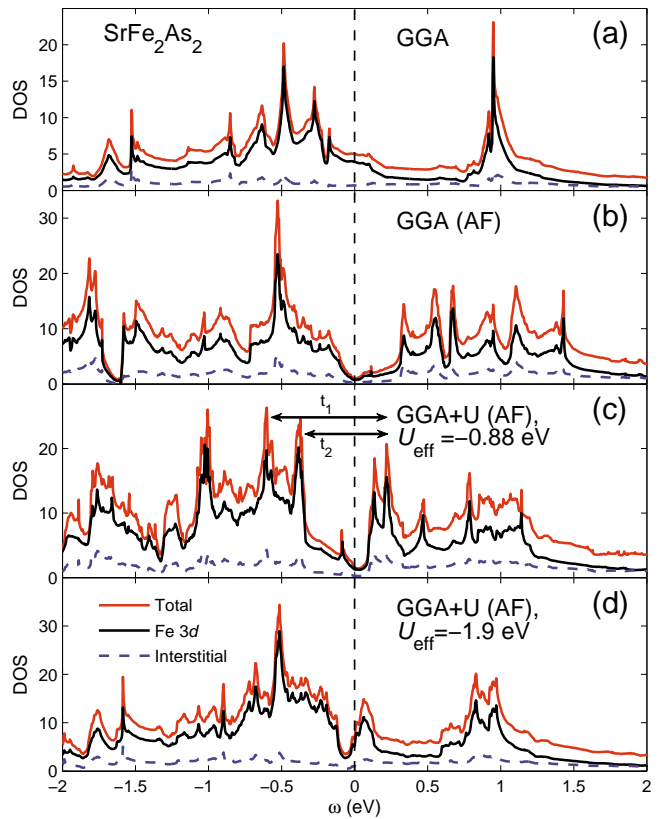}
\caption{\label{fig:Sr_dos}(Color online) DOS of {\Sr} for different values of $U_{\rm eff}$. $\omega=0$ corresponds to the Fermi energy. As is well known, the Fe~$3d$ subshell dominates the DOS around the Fermi energy. The contributions from Sr and As, and the other Fe orbitals are small. The magnetic moments from top to bottom are $m=0$, $m=1.98\:\muB$, $m=1.30\:\muB$, and $m=0.43\:\muB$, respectively. See text for a discussion of the transitions labelled by t$_1$ and t$_2$.}
\end{figure}

For the iron pnictides, DFT calculations are known to strongly
overestimate the magnetic moments on the Fe atoms in the SDW state
($\approx\!2.0 \muB$ with GGA(AF) compared to $0.4
\muB$~\cite{Cruz2008}$-0.6\muB$~\cite{Qureshi2010} for {\LaAs}). As
will be discussed in the following, this results in a too large SDW
gap which in turn shifts the SDW features in the optical properties
to erroneously high energies. In order to reproduce correctly the
SDW gap, we employ GGA+U with $U_{\rm eff}$
negative~\cite{Nakamura2009,Yi2009} to suppress the overestimated
magnetic moment from GGA(AF). On the one hand, the negative $U_{\rm
eff}$ can be viewed as a result of an overscreening of the on-site
Coulomb interaction which leads to a smaller value of U compared to
the on-site exchange interaction J or a strong Holstein-like
electron-phonon coupling which gives a frequency-dependent negative
contribution to the on-site Coulomb interaction. On the other hand,
similar to the application of positive U at the mean-field level in
GGA+U calculations to reproduce the correlated gap in
Mott insulators, the negative $U_{\rm eff}$ can be understood as a
simple way to suppress the tendency to
magnetism within the mean-field approximation and mimic the effect of quantum fluctuations ignored in DFT. This can be read off from the correction to
the GGA functional in Eq.~(\ref{eq:u_correction}) which energetically
favors $n_{m\sigma}=1/2$ at negative $U_{\rm eff}$, i.e. the paramagnetic case, and conversely
penalizes (spin-projected) integer occupations, i.e. the fully polarized states.
Note that the exact form of the GGA+U functional and thus the preference of
low-spin or high-spin states depends on the double counting correction.\cite{Ylvisaker2009} A comparison of
the role of U and J with their model counterparts is therefore not generally valid
but needs to take into account the double counting correction under consideration.
Also, the magnetic moment needs to be further reduced
below its experimental value due to the RPA employed in our calculations of optical conductivities, as discussed later. Therefore, the negative $U_{\rm eff}$ used in this paper is not comparable
to realistic values of U and J for these systems due to the approximations we employed here.

\begin{figure*}[tb]
\includegraphics[width=\textwidth]{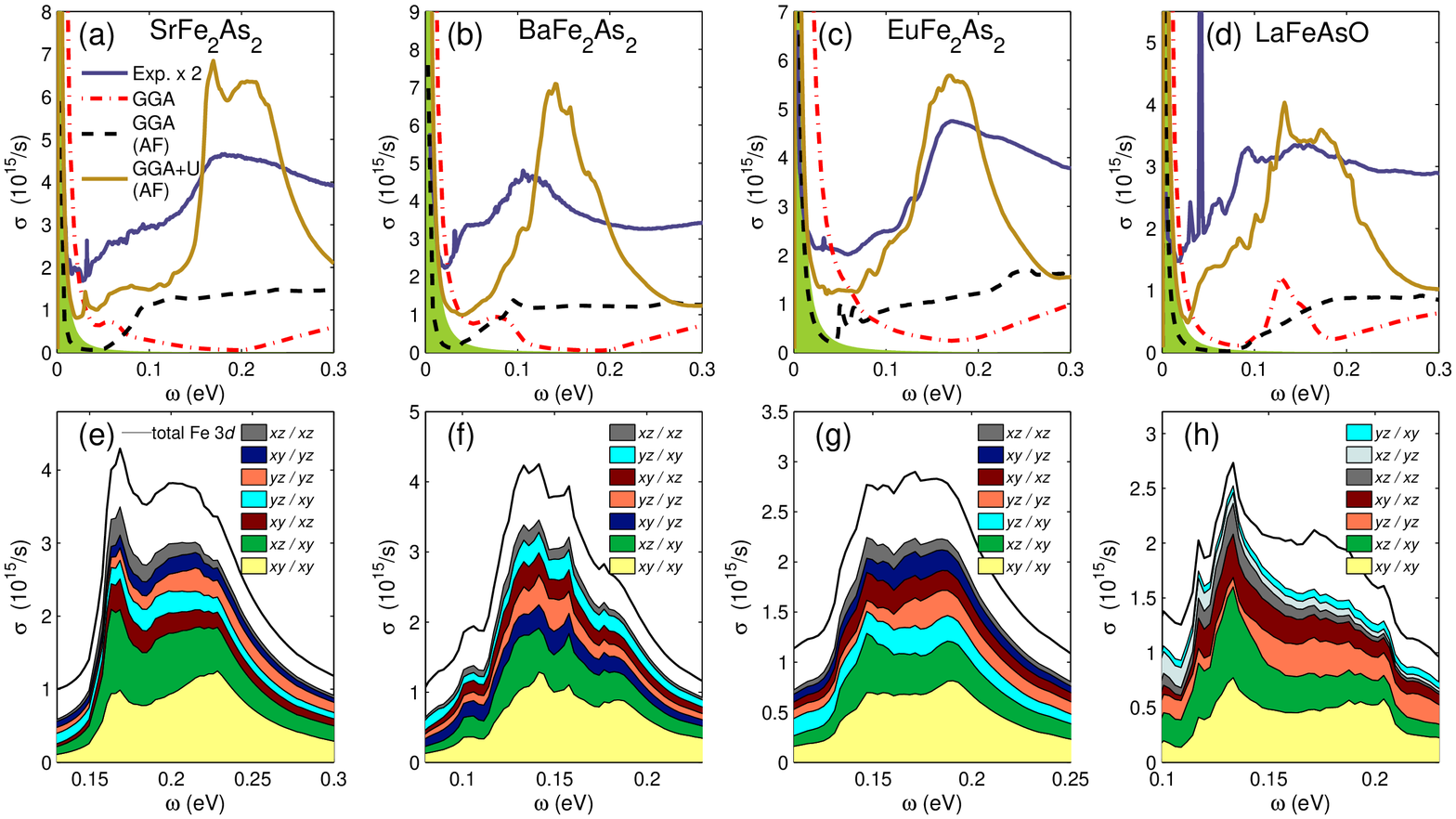}%
\caption{(Color online) Real part of the optical conductivity of
  different iron pnictides in the SDW state, for
  $U_{\rm eff}=-1.9$ eV: (a),(e) {\Sr} (experimental data taken from
  Ref.~\onlinecite{Hu2008}), (b),(f)
  {\Ba}~(Ref.~\onlinecite{Hu2008}), (c),(g) {\Eu}
  (Ref.~\onlinecite{Wu2010}), and (d),(h)
  {\LaAs}~(Ref.~\onlinecite{Chen2010}). No
  (empirical) Lorentz broadening has been applied; in order to
  facilitate the comparison with the damped/broadened experimental
  data, the experimental results have been rescaled by a factor of
  two. The orbital character
  contributions are sorted by their contribution to the total
  conductivity, integrated over the energy range shown in the
  respective plot, with the largest contribution at the bottom (for
  clarity, only the first seven contributions are displayed; smaller
  contributions have been omitted).}
\label{fig:conductivity_lowf}
\end{figure*}

Fig. \ref{fig:Sr_dos} shows the evolution of the density of states for
different values of $U_{\rm eff}$ in {\Sr}. Compared to the
non-spin-polarized GGA DOS, a suppression of the DOS around $E_{\rm
  F}$ is clearly visible in the antiferromagnetic calculations.
For the parameter regimes shown in Fig. \ref{fig:Sr_dos}, only partial
opening of the gap is observed as some DOS persists around $E_{\rm
  F}$; full opening of the gap occurs at positive values of $U_{\rm
  eff}$, {\it i.e.} for sufficiently strong Hubbard U. In contrast,
for pronounced negative $U_{\rm eff}$ the magnetic moment is reduced
and the SDW gap narrows accordingly. For $U_{\rm eff}=-1.9$ eV and
$m=0.43\: \muB$, the energy gap is $\Delta\approx 0.17$ eV. Almost
over the whole energy range shown in the figure, the DOS is strongly
distorted and shifted by the inclusion of the antiferromagnetic
order compared to the non-spin-polarized GGA DOS. However, the
strong suppression of the magnetic moments for $U_{\rm eff}=-1.9$ eV
renders the DOS almost equal to the GGA DOS again, except in the
region around the Fermi energy. Thus, the optical properties can be
expected to be close to the GGA case in the higher energy regions of
the spectrum.

\subsection{Optical conductivity}

As for the in-plane optical conductivity of the iron pnictides in the
SDW state, experimental investigations on single
crystals~\cite{Wu2010,Hu2008,Chen2010,Lucarelli2010} consistently report a number of
common features, although the exact location of the features depends
on the material under investigation.  These features are: metallic
behaviour, {\it i.e.} the presence of a Drude-like conductivity at low
frequencies ($\alt 100$ cm$^{-1}$), a sharp peak at the SDW gap
frequency ($\approx 1000$ cm$^{-1}$--1500 cm$^{-1}$), and a broad,
less pronounced peak in the midinfrared region ($\approx 5000$
cm$^{-1}$--6000 cm$^{-1}$) which almost doesn't depend on temperature
and is present also above $T_{\rm SDW}$. The peaks are associated with
a suppression of the spectral weight at lower energies (below $\approx
1000$ cm$^{-1}$ for the SDW peak, below $\approx 5000$ cm$^{-1}$ for
the high-energy peak) which leads to a spectral weight transfer from
lower to higher energies. Note that in the normal state, the
tetragonal symmetry leaves only two independent components in the
dielectric tensor and thus the conductivity is defined as
$\sigma_{aa}=\sigma_{bb}$ for the Fe in-plane directions $a$ and $b$,
and $\sigma_{cc}$ for the out-of-plane direction $c$ perpendicular to
$a$ and $b$. In the SDW state, the stripe-like AF order introduces an
anisotropy which lifts the degeneracy of the two in-plane
components. However, no substantial anisotropy is found
experimentally.

\begin{figure}
\includegraphics[width=\columnwidth]{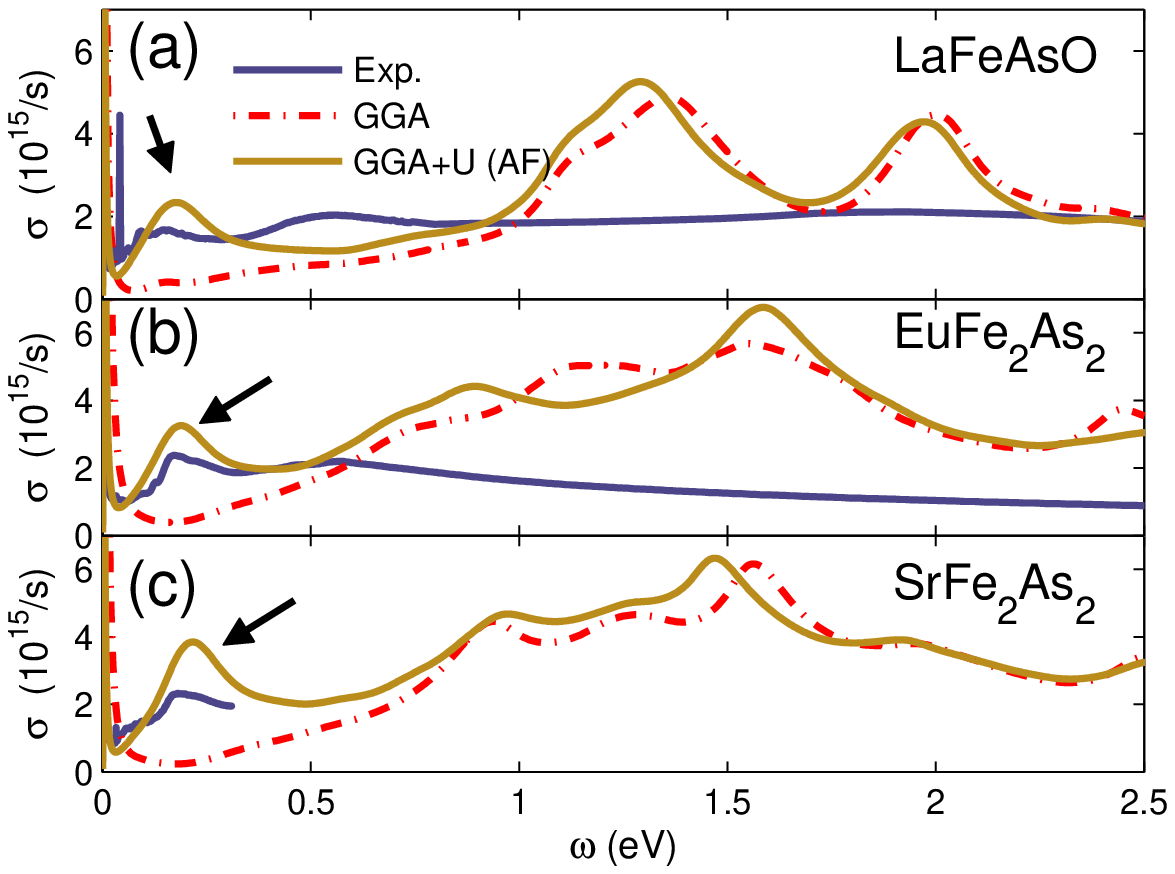}%
\caption{\label{fig:conductivity_highf}(Color online) Optical
  conductivity in the higher energy region for (a) {\LaAs}, (b)
  {\Ba}, and (c) {\Sr}. Same experimental references and values for $U_{\rm eff}$ as
  in Fig. \ref{fig:conductivity_lowf}. The arrows indicate the position of the respective SDW peak.}
\end{figure}

\begin{figure}
\includegraphics[width=\columnwidth]{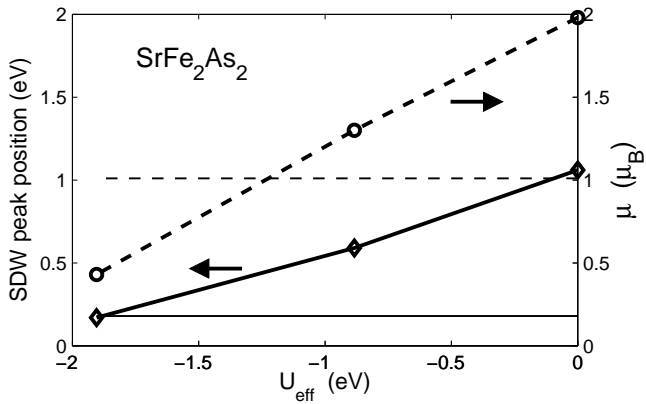}%
\caption{\label{fig:Sr_params}Dependence of the SDW
  peak position in the optical conductivity and
  the magnetic moment on the value of $U_{\rm eff}$ for {\Sr}. The
  thin lines indicate experimental values: $\mu_{\rm exp}=1.01
  \muB$~\cite{Kaneko2008, Jesche2008}, SDW$_{\rm exp}=0.18$
  eV~\cite{Hu2008}}.
\end{figure}

In Fig. \ref{fig:conductivity_lowf} we present an overview of the
in-plane optical conductivity in the low-frequency region as obtained
by our calculations. Figs. \ref{fig:conductivity_lowf} (a)--(d) show a comparison of the
experimental results with the calculation methods introduced in the
previous section; the GGA+U(AF) results exhibit the SDW peak as the
most prominent feature, located at the SDW gap frequency. For all
compounds, the SDW peak emerges at the experimentally determined
frequency only for a suitable negative value of $U_{\rm eff}$, and otherwise moves to higher
frequencies (see Fig. \ref{fig:Sr_params} for {\Sr}). The value $U_{\rm eff}=-1.9$ eV was approximately
determined by demanding a correct SDW peak position. Fig. \ref{fig:conductivity_highf} displays the optical conductivity
for {\LaAs}, {\Ba}, and {\Sr} over a larger energy range and includes some
damping. As can be seen there, the GGA results without SDW order don't
show any significant peak in the optical conductivity up to $\approx
1$ eV. This can be readily read off from the density of states, where
the GGA DOS is essentially depleted up to 1 eV above $E_{\rm F}$. At higher energies, there is an obvious
disagreement between the DFT results and the experimental data, for both GGA+U(AF) and GGA.
Whereas we already mentioned the midinfrared peak around 0.6 eV, the experimental data
basically doesn't show any structure above 1 eV. This could be described if quantum fluctuations are treated properly
and consequently a frequency-dependent self-energy is involved. However, as the focus of this work is the analysis
of the features induced by the antiferromagnetic ordering, we concentrate on the low energy features as
indicated by the arrows.

As for the anisotropy due to the stripe-type SDW symmetry-breaking, we
find that it is strongly present in the GGA(AF) calculation ({\it
  i.e.} for $U_{\rm eff}=0$) and for moderately negative values
$U_{\rm eff}\agt -1.5$ eV, where $\sigma_{bb}$ -- the conductivity
along the FM axis -- shows a peak at significantly higher
frequencies than $\sigma_{aa}$. In fact, this anisotropy can be
mapped to the DOS in Fig. \ref{fig:Sr_dos}, where t$_1$ denotes the
transition(s) which dominate the peak in $\sigma_{bb}$ and t$_2$ the
ones dominating $\sigma_{aa}$. As $U_{\rm eff}$ is decreased, the
anisotropy is gradually suppressed and for the regime shown in Fig.
\ref{fig:conductivity_lowf}, almost no anisotropy is present anymore
in {\Sr}, {\Ba} and {\Eu}, in accordance with the experimental
observations. This can be understood by the fact that while d$_{xz}$
and d$_{yz}$ are degenerate in the high temperature tetragonal
phase, below the SDW transition temperature, different band
splittings and band shiftings are produced by different
magnetization and occupation number on these two orbitals.
Therefore, the excitations along $x$ and $y$ direction become
inequivalent. As $U_{\rm eff}$ is decreased, the magnetization and
consequently the symmetry-breaking between d$_{xz}$ and d$_{yz}$
orbitals are suppressed, leading to smaller differences between
d$_{xz}$ and d$_{yz}$ orbitals and accordingly in the excitations
along $x$ and $y$ direction. An anisotropy can still be seen in the
weak double peak structure of {\LaAs}~ and {\Sr} in Figs.
\ref{fig:conductivity_lowf} (d) and (h), respectively (a) and (e)  (the conductivity shown in
the plots is given by $\sigma=(\sigma_{aa}+\sigma_{bb})/2$): e.g. in {\LaAs}, 
the d$_{xy}$ contribution is of roughly equal size in both sub-peaks, whereas the 
d$_{xz}$ contribution is mainly present in the first sub-peak and the d$_{yz}$
contribution in the second one. Consequently, the second sub-peak is dominated
by $\sigma_{bb}$, {\it i.e.} by the conductivity along the FM direction. We want to point out that also this sub-peak is directly related
to the AF order as no peak is present in a purely ferromagnetic
calculation (with the same absolute value of the magnetic moments).

The observed midinfrared peak at $\approx\!\!0.6$ eV, which is visible
in Fig. \ref{fig:conductivity_highf} for {\LaAs} and {\Ba},
respectively, is absent in all calculations, including the
non-spin-polarized GGA calculations. This is in contrast to the
experimental data and needs to be investigated further.

Surprisingly, the calculated optical spectrum of {\Sr} shows an
experimentally observed low-frequency feature at 31~meV (see Fig. \ref{fig:conductivity_lowf}(a)) that has so
far been attributed to phononic excitations: while there is no
discussion in the respective experimental work
Ref.~\onlinecite{Hu2008} for {\Sr}, for {\Eu} it is argued in
Ref.~\onlinecite{Wu2009a} that the phononic mode -- there at 32~meV
-- exhibits a coupling to the electronic background only for high
temperatures while for low temperatures the Lorentz line shapes
indicates a purely phononic, non-interacting excitation. However,
our results show that this peak can also be obtained from the
electronic band structure alone in {\Sr}.

In Figs. \ref{fig:conductivity_lowf} (e)--(h), the
decomposition of the SDW peak into Fe~$3d$ orbital characters of the
initial and final states is shown. The orbital resolved optical conductivity
reads
\begin{equation}\begin{split}
\sigma^{m_i,m_f}(\omega)=&\frac{\hbar^2 e^2}{4 \pi^2 m^2 \omega} \sum_{v,c} \int_{\vec{k}} A_{c,k}^{m_f} \; |\bra{c,\vec{k}}p\ket{v,\vec{k}}|^2 \\
& \times A_{v,k}^{m_i} \;
\delta(E_{c,\vec{k}}-E_{v,\vec{k}}-\hbar\omega),
\label{eq:sigma_ch_resolved}
\end{split}\end{equation}
where $A_{n,k}^{m}$ formally denotes the relative weight of the Fe
orbitals with magnetic quantum number $m$ in the KS orbital
$\ket{n,\vec{k}}$. Note that the momentum operators only couple
states with different parity, reflected by the well known selection
rule $\Delta l=l_f-l_i=\pm 1$ for dipolar transitions. Dipolar
transitions among Fe~$3d$ states are therefore forbidden in the
atomic limit. For the iron pnictides, the states around the Fermi
energy are Fe~$3d$ dominated with some As $4p$ contribution. This
hybridisation allows for transitions between initial and final
states which are both Fe~$3d$ dominated but owe their finite
transition strength to ${\rm
  Fe} \:3d \leftrightarrow {\rm As}\: 4p$ transitions. This is taken
into account in Eq.~(\ref{eq:sigma_ch_resolved}) where the full
wavefunction $\ket{n,\vec{k}}$ is used for the calculation of the
matrix element but the resulting dielectric function is projected on
the Fe~$3d$ subspace. This projection in particular omits the
contribution from the interstitial region which has a considerable DOS
in the considered energy range but cannot be assigned an orbital
character. However, as can be seen from the comparison of the SDW
peaks in Figs. \ref{fig:conductivity_lowf} (a)--(d) -- which
show the total optical conductivity -- to the close-up in (e)--(h) -- which only show the Fe~$3d$ contribution --, the Fe~$3d$
part resembles very well the structure of the total
conductivity. Therefore the contributions from the interstitial region
as well as from the other atoms are neglected for the orbital
character analysis.

In the iron pnictides, the Fermi surface is crossed by multiple
bands. Consequently, also the optical properties have multiband nature
and one can expect contributions from several orbital characters. This
is confirmed by our calculations where we observe no overly dominating
character component in any part of the spectrum. The SDW peak
structure differs from the rest of the spectrum in that it mainly
contains $t_{2g}$ character components; all $t_{2g}$ components are
larger than any $e_g$ component.

\begin{table}[h!]
  \caption{\label{tab:results}Optical parameters of the investigated compounds as obtained from the different calculation methods. $\omega_p^{a(b)}$ is the plasma frequency in $a$($b$)-direction.}
\begin{ruledtabular}
\begin{tabular}{llllll}
Compound & Calc. & $\omega_p^a$ [eV] & $\omega_p^b$ [eV] &\# bands & $m$ [$\muB$]\\
& & & & at $E_{\rm F}$ &\\
\hline
{\LaAs} & GGA & 2.25 & 2.25 & 5 & 0\\
& GGA(AF) & 0.87 & 0.82 & 2 & 1.98\\
\vspace{-1ex} & GGA+U(AF), & 1.74 & 1.17 & 2 & 0.29\\
& $U_{\rm eff}=-1.9$eV & & & &\\
{\Ba} & GGA & 2.62 & 2.62 & 5 & 0\\
& GGA(AF) & 0.62 & 0.65 & 2 & 1.98\\
\vspace{-1ex} & GGA+U(AF), & 1.66 & 1.53 & 3 & 0.30\\
& $U_{\rm eff}=-1.9$eV & & & &\\
{\Sr} & GGA & 2.79 & 2.79 & 5 & 0\\
& GGA(AF) & 0.64 & 0.64 & 2 & 1.98\\
\vspace{-1ex} & GGA+U(AF) & 1.50 & 1.27 & 2 & 0.43 \\
& $U_{\rm eff}=-1.9$eV & & & & \\
{\Eu} & GGA & 2.96 & 2.96 & 5 & 0\\
& GGA(AF) & 0.84 & 0.99 & 2 & 1.76\\
\vspace{-1ex} & GGA+U(AF) & 1.95 & 2.04 & 3 & 0.34\\
& $U_{\rm eff}=-1.9$eV & & & &\\
\end{tabular}
\end{ruledtabular}
\end{table}

Finally, we analyze the low-energy Drude region of the spectrum as
characterized by the plasma frequency. Since the SDW gap opening is
only partial, the Drude peak is still present -- expressed by a
finite plasma frequency --  both experimentally and  in our
calculations. The ratio of the kinetic energies which equals the
ratio of the squares of the plasma frequencies,
$K_{\rm exp}/K_{\rm band}=(\omega_p^{\rm exp})^2/(\omega_p^{\rm
  band})^2$, is commonly taken as a measure for the renormalization
effect from the electronic correlations compared to band structure
calculations. As given in Table \ref{tab:results}, the
non-spin-polarized GGA value for, {\it e.g.}, {\Sr} is
$\omega_p\approx 2.79$ eV, whereas the experimental value in the
normal state is $\omega_p\approx 1.7$ eV at 300 K,~\cite{Hu2008}
yielding $K_{\rm exp}/K_{\rm band}\approx 0.37$. In the SDW state, the experimental plasma frequency is strongly
reduced due to the removal of itinerant carriers from the Fermi
surface by the opening of the SDW gap, to $\omega_p\approx 0.59$ eV at
10 K.~\cite{Hu2008} Likewise, in our calculations, the inclusion of
the SDW order reduces the number of bands crossing the Fermi surface,
which also significantly reduces the plasma frequencies. The
renormalization due to the correlations still persists, though: with
$K_{\rm exp}/K_{\rm band}\approx 0.15$ at $U_{\rm eff}=-1.9$ eV it is
even more pronounced in the SDW state.

In Table \ref{tab:results} as well as Fig. \ref{fig:Sr_params}, one
also notices that with our choices of $U_{\rm eff}$ in the GGA+U(AF)
calculations, experimental SDW peak positions are
well reproduced, while the calculated magnetic moments are all
smaller than the corresponding experimental values, for example,
the calculated magnetic moment for {\Sr} is $0.43\:\mu_B$ while it is
$1.01\:\mu_B$ from neutron diffraction experiments~\cite{Kaneko2008,
Jesche2008}. The discrepancy is due to the fact that the
employed RPA scheme neglects correlation effects, like interactions
between electrons and holes which may reduce the excitation energy
and would therefore shift the SDW peak towards lower frequencies. In
order to account for such a shift induced by  correlation effects,
a smaller value of the magnetic moment compared to the experimental one has
to be used, which similarly shifts the SDW peak towards a  lower
position. For {\LaAs}, the discrepancy between the calculated magnetic moment, 
with which the experimentally observed SDW peak position is reproduced, and the 
experimental one is less
pronounced, indicating less correlation strength in {\LaAs}~than in
the other three compounds which is consistent with the DFT downfolding
results of Ref.~\onlinecite{Miyake2010}.

For all compounds except {\Eu}, the plasma frequencies and thus the dc
conductivities exhibit a notable anisotropy between the $a$ and $b$
axis. Interestingly, the anisotropy develops differently in the 1111
compound and the 122 compounds where the 1111 compound always
features a higher conductivity along the $a$ direction. In contrast,
$\sigma^{b}>\sigma^{a}$ in the 122 compounds for $U_{\rm eff}=0$
which is expected because of the larger lattice constant along the
$a$ axis and the orientation of the SDW vector along $a$.
Unexpectedly, this no longer holds for negative $U_{\rm eff}$ where
we find $\sigma^{a}>\sigma^{b}$ also for the 122 compounds. This is
in agreement with recent experiments on underdoped {\Ba} where the
magnetic domains were (partially) detwinned in a magnetic field in
order to reveal the in-plane anisotropy in the
resistivity.~\cite{Chu2009}

\section{Conclusions}

In summary we have demonstrated that DFT is capable to reproduce a
number of features associated with the SDW state in the iron
pnictides. However, this comes at the cost of a negative $U_{\rm
eff}$ in the context of GGA+U calculations. The negative U may be
understood as a way of simulating a strong screening of the Coulomb
interaction and a large electron-phonon
coupling~\cite{Liu2009,Shirage2009,Rahlenbeck2009,Choi2010,Kordyuk2010,Boeri2010}.
Also it can be viewed as a route to mimic the effects of quantum
fluctuations at the mean-field level and therefore as a driving
force for suppressing the overestimated Fe magnetic moments obtained
from GGA(AF).  Concerning the size of the magnetic moments and thus
the position of the SDW peak, the negative $U_{\rm eff}$ therefore
needs to be interpreted as a fitting parameter. However, the
agreement with the experimentally observed isotropy of the in-plane
conductivity and the appearance of the low-frequency
phononic-electronic excitation at exactly the right frequency
indicate the reliability of the approach. Therefore, we conclude
that the used LDA+U framework with negative $U_{\rm eff}$ mainly
reduces the magnetic moment but doesn't distort the overall band
structure too seriously. This suggests that this method is
surprisingly well suited for the description of the SDW state in the
iron pnictides.

{\it Acknowledgements}.- We would like to thank M. Dressel,
N. Bari{\v{s}}i\'c, N. Drichko, J. Fink and K. Foyevtsova for useful
discussions. We gratefully acknowledge financial support from the
Deutsche Forschungsgemeinschaft through the SPP 1458 program and from
the Helmholtz Association through HA216/EMMI.

%\bibliography{p1}

\end{document}